# Reconfigurable Momentum-space vectorial lasing enabled by Quasi-BIC

Hongyu Yuan, Zimeng Zeng, Jiayao Liu, Zhuoyang Li, Xiaolin Wang, Zelong He and Zhaona Wang[*]

Key Laboratory of Multiscale Spin Physics (Ministry of Education), Applied Optics Beijing Area Major Laboratory, School of Physics and Astronomy, Beijing Normal University, Beijing 100875, China

[*]*Contact author: zhnwang@bnu.edu.cn*

**Abstract**

Bound states in the continuum (BICs) have enabled lasers with rich momentum-space textures. However, the output patterns of quasi-BIC lasers remain largely static and confined to a few geometries. Here, a reconfigurable momentum-space vectorial laser was proposed based on two-dimensional photonic crystal. By selectively exciting quasi-BIC modes, we identify the geometric asymmetry factors favoring single BIC, dual-BIC, and radiative mode with BIC operation. This approach yields vectorial lasing with characteristic patterns lasing in momentum space of bidirectional double lobes (BDL), radially polarized ring with BDL, azimuthally polarized ring with BDL, and linearly polarized spot with BDL. Importantly, reversible switching between a single donut and a donut with BDL was achieved in the same device by varying the pump energy density. Our work establishes a compact, versatile platform for reconfigurable vectorial lasers, with potential applications in tunable optical tweezers, super-resolution imaging, and on-chip optical interconnects.

**Introduction**

The precise manipulation of vectorial optical fields in momentum space enables the synergistic engineering of radiation intensity distribution and polarization orientation along each wavevector direction[1-4]. This capability offers significant potential for applications in optical communications[5], optical tweezers[6], and super-resolution imaging[7]. Over the past decade, significant efforts have been devoted to generating and controlling such vectorial optical fields[8,9], primarily by either passively shaping incident laser beams using external diffractive elements[10] or directly emitting vectorial cavity modes from lasers[11,12]. For instance, vectorial fields with tailored polarization distributions have been generated by integrating dielectric metasurfaces onto vertical-cavity surface-emitting lasers[13]. However, such passive wavefront shaping relies on phase modulation during propagation, which inevitably introduces additional efficiency losses. In contrast, vectorial fields originating directly from cavity modes offer higher efficiency and superior stability[14,15], as their polarization properties are inherently governed by cavity symmetry and mode selection. A notable example is the orbital angular momentum microlaser reported by Miao et al.[16], in which whispering-gallery modes are tailored to directly emit radially polarized vectorial vortex beams, highlighting the potential of on-chip vectorial light sources.

Beyond conventional cavity, photonic crystals (PCs) provide a powerful platform to design cavity mode and manipulate the vector light field[17-19]. Among these cavity modes, bound states in the continuum (BICs) are a distinct class of non-radiative eigenmodes that manifest as topological polarization singularities in momentum space[20,21]. Under weak perturbations, they transform into quasi-BICs with high quality ($Q$) factors while typically preserving the vectorial polarization nature[22,23]. This unique combination makes them particularly attractive for constructing vectorial optical fields with rich momentum-space textures, thereby expanding the dimensions of light field manipulation[24,25]. Leveraging these advantages, quasi-BIC lasers have been widely explored for low-threshold operation[26-28] and for generating vectorial emission with tailored polarization distributions[29-31].

To date, a variety of radiation patterns have been demonstrated in quasi-BIC lasers based on one-dimensional (1D) and two-dimensional (2D) PCs, including double-lobe patterns[32] or ring-shaped emissions[33-36] in momentum space. Notably, Song et al. recently demonstrated a metalaser in which the interplay between local and nonlocal responses enables arbitrary wavefront control via spatially varying geometric phases[37], showcasing the potential of quasi-BIC lasers for programmable emissions in momentum space. Unlike conventional approaches that primarily modulate intensity, quasi-BICs provide simultaneous control over phase and polarization degrees of freedom by engineering the interference of leaky channels, a capability essential for synthesizing vectorial fields with tailored wavefronts and polarization topologies[38,39]. Despite these theoretical possibilities, experimental radiation patterns remain limited in both diversity and reconfigurability. Most quasi-BIC lasing fields are intrinsically static, with their emission profiles predetermined by fixed cavity geometries. This lack of dynamic controllability poses a significant barrier for applications requiring adaptive wavefronts or reconfigurable polarization responses in multifunctional photonic integrations. As a step toward addressing this challenge, we recently demonstrated dynamic switching between quasi-BIC and Bragg resonance modes in 1D PCs by controlling the pump polarization and filling factor[40], revealing the critical role of fine structural parameters in tailoring light fields. These results suggest that deliberate design of quasi-BIC cavity geometries, enabling mode competition and coupling, is essential not only for achieving customized vectorial emissions but also for generating complex vectorial fields with on-demand phase and polarization distributions.

Here, we propose and experimentally demonstrate a reconfigurable momentum-space vectorial laser based on a 2D square-lattice PC with an engineered geometric asymmetry factor ($\alpha$) of the unit pillars. The cavity features elliptical pillars with the dimension fixed along $y$ and varied along $x$, thereby introducing a geometric asymmetry that breaks the lattice symmetry from $C_{4v}$ to $C_{2v}$. This asymmetry factor controls the $Q$ factors of BICs and radiative modes, enabling precise manipulation of their mutual coupling and competition. By tuning $\alpha$, we find three distinct operational regimes of single-BIC, dual-BIC, and radiative mode with BIC. Specifically, at $\alpha = -0.02$, single quasi-BIC lasing yields a bidirectional double lobes (BDL) pattern in momentum space. At $\alpha = -0.04$ and $-0.18$, the two quasi-BIC lasing modes produce radially or azimuthally polarized rings superimposed on BDL, respectively. At $\alpha = -0.23$, a hybrid lasing regime combining quasi-BIC and radiative mode is achieved, featuring a linearly polarized spot coexisting with the BDL. Remarkably, in the dual-BIC regime, switching between a single donut and a donut with BDL is demonstrated by simply varying the pump energy density, indicating dynamic reconfigurability of momentum-space textures within a fixed cavity geometry. These findings demonstrate that deliberate symmetry engineering provides a robust pathway to achieve both static pattern diversity and dynamic reconfigurability in quasi-BIC vectorial lasers, holding promise for applications such as high-efficiency particle manipulation and laser processing.

**Theoretical design and experimental Results**

**Tuning principle of momentum-space radiation patterns in 2D PC microcavity**

The designed 2D PC-waveguide coupled microcavity doped with Rhodamine 6G (R6G) consists of a uniform waveguide layer with a thickness of 1.75 ± 0.05 μm and a 2D square PC layer comprising elliptical polymer pillars arranged with lattice constants of $\Lambda_x = \Lambda_y = 375$ nm (Fig. 1(a)). The pillar axis lengths along the $x$ and $y$ direction are denoted as $w_x$ and $w_y$ (inset of Fig. 1(a)), respectively. To introduce the geometric asymmetry factor $\alpha = (w_x - w_y)/(w_x + w_y)$, we fix $w_y = 187.5$ nm and systematically varying $w_x$. This coupled structure supports transverse electric (TE) and transverse magnetic (TM) polarized eigenmodes, labeled as $TE_{l,p}^m$ and $TM_{l,p}^m$, respectively (Fig. S1 (a)), where $l$ and $p$ are the diffraction orders along the $x$ and $y$ directions, and $m$ is the mode order in the vertical direction[41]. In the absence of coupling, the photonic band structure of each eigenmode can be derived analytically using the generalized Bragg condition (Fig. S1 (b) and (c)). For $m = 1$, the TE eigenmodes consists of two pairs of Bloch modes arising from $x$-direction ($l = \pm 1$), and from $y$-direction ($p = \pm 1$), respectively. Similarly, the corresponding TM band branches follow the same rule and are named as $TM^1_{-1,0}$, $TM^1_{+1,0}$, $TM^1_{0,-1}$ and $TM^1_{0,+1}$. Under the influence of mode coupling, band anti-crossing occurs near the Bragg resonance[42,43]. Taking the 2D quasi-BIC microcavity with $\alpha = -0.02$ as an example, the dispersion curves of the eigenmodes along the Γ–X direction are simulated using a three-dimensional finite element model (Fig. 1 (b)). Based on their Γ-point $Q$ factors and symmetry of the localized electromagnetic field (Fig. S2(a)), the eigenmodes are classified as non-radiative dark modes (BICs), i.e., $TE_{d1}$, $TE_{d2}$, $TM_{d1}$, and $TM_{d2}$, and radiative bright modes, i.e., $TE_{b1}$, $TE_{b2}$, $TM_{b1}$, and $TM_{b2}$.

Furthermore, the tuning behavior of the asymmetry factor $\alpha$ on the $Q$ factors of various resonance modes was systematically investigated in the momentum space for across varying $\alpha$ from $-0.33$ to $0.05$ (Fig. 1 (c) and Fig. S2 (b-e)). Across the entire tuning range, all BICs maintain high $Q$ factors exceeding $10^9$, a property protected by the $C_{2v}$ symmetry subgroup[44,45]. Specially, $TE_{d1}$-derived quasi-BIC exhibits low radiation loss across all $\alpha$ values when the in-plane wavevector deviates from the Γ point, indicating strong fabrication tolerance and weak loss through the off-axis radiation channel. These characteristics give this mode an advantage in achieving lasing over other quasi-BIC modes corresponding to $TE_{d2}$, $TM_{d1}$, and $TM_{d2}$, enabling the $TE_{d1}$-derived quasi-BIC lasing. Importantly, the $Q$ factors of the $TE_{d2}$ and $TM_{d1}$ modes are significantly enlarged by over three orders of magnitude at the special $\alpha$ values of $\alpha = -0.18$ ($TE_{d2}$) and $\alpha = -0.04$ ($TM_{d1}$), indicating stronger localization effect and smaller emission loss through the vertical channel. This enlarged Q factors allow their corresponding quasi-BICs with sufficiently low radiation loss to also support lasing at the special structure parameters. This behavior, together with the consistently favorable performance of the $TE_{d1}$-derived quasi-BIC, establishes a mechanism for dual quasi-BIC lasing. For radiative modes, the $Q$ factors increase monotonically as $\alpha$ decreases, indicating progressively reduced radiation losses (Fig. 1(c)). At $\alpha = -0.23$, the $Q$ factor of the radiative mode $TE_{b2}$ reaches $8 \times 10^4$, providing favorable conditions for Bragg resonance lasing[40]. This behavior, combined with the consistently low radiation loss of the TE$_{d1}$-derived quasi-BIC, enables a distinct regime in which a radiative mode and a quasi-BIC can simultaneously support lasing. Through the precise controllability over $\alpha$, three distinct regimes can be realized, including single-BIC, dual-BIC, and radiative mode with BIC. Such controllability over the radiation states underpins the diverse momentum-space lasing patterns.

As illustrated in Fig. 1(d), the schematic momentum-space radiation patterns corresponding to the three regimes described above reveal distinct far-field characteristics determined by the number and type of excited quasi-BICs or radiative modes. For single BIC regime ($\alpha = -0.02$), only TE$_{d1}$-derived quasi-BIC is excited, yielding a BDL pattern which consists of two bright lines parallel to the $k_x$ axis and two parallel to the $k_y$ axis, with vanishing intensity along both $k_y = 0$ and $k_x = 0$. This BDL pattern remains present across all $\alpha$ values, serving as a basic lasing mode. For dual-BIC regime, the additional quasi-BICs derived from TM$_{d1}$ ($\alpha = -0.04$) and TE$_{d2}$ ($\alpha = -0.18$) become excitable alongside the TE$_{d1}$-derived quasi-BIC, resulting in more complex momentum-space textures. Each of these additional quasi-BICs manifests as a ring-like structure centered around the $\Gamma$-point, superimposed on the BDL pattern. Although TM$_{d1}$- and TE$_{d2}$-derived quasi-BICs show similar ring-shaped intensity profiles, they differ in far-field polarization due to their distinct symmetries. For radiative mode with BIC regime ($\alpha = -0.23$), the radiative mode TE$_{b2}$ is excited together with TE$_{d1}$-derived quasi-BIC, introducing a spot-like pattern at the $\Gamma$ point superimposed on the BDL pattern. These distinct momentum-space patterns establish the foundation for further analysis of polarization behavior in the three regimes.

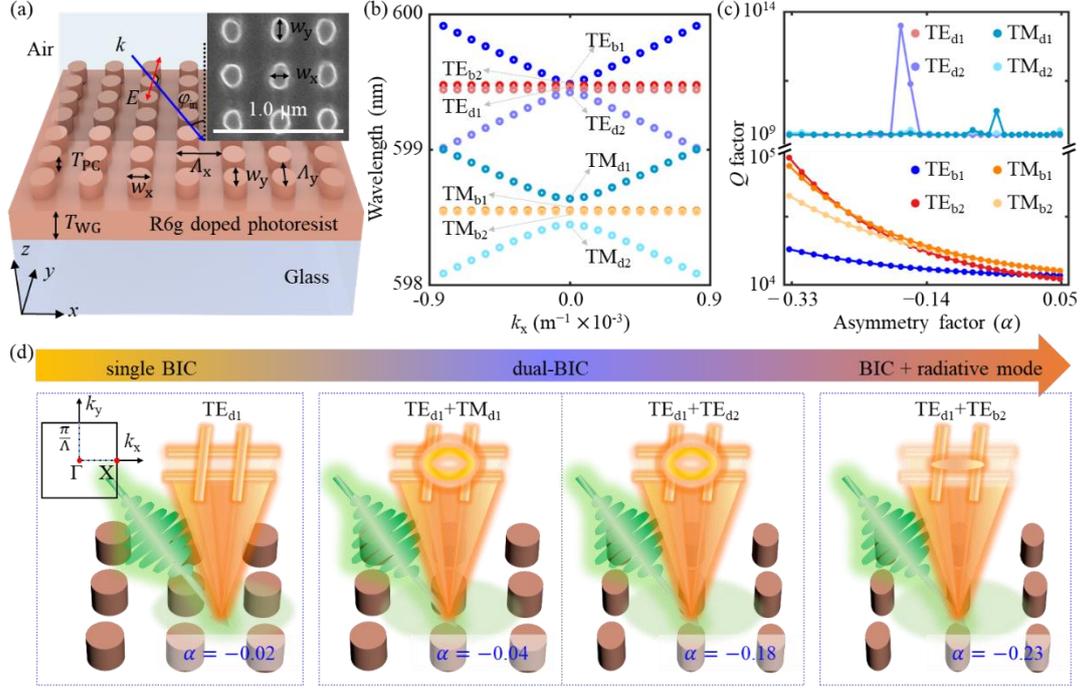

Fig. 1. Designed 2D PC-waveguide coupled microcavity and its optical mode characteristics. (a) Schematic of the structure with corresponding SEM image in the inset. (b) Simulated dispersion curves along Γ–X at $\alpha = -0.02$. (c) $Q$ factors of resonant modes versus $\alpha$. (d) Momentum-space radiation patterns illustrating single BIC, dual-BIC, and radiative mode with BIC regimes.

**Designing four types of vectorial lasing in momentum-space**

The The polarization characteristics are further comprehensively analyzed for the selected $\alpha$ values to present the emission property of the three distinct regimes (Fig. 2). In the single BIC regime ($\alpha = -0.02$ in Fig. 2(a)), the BIC $TE_{d1}$ (pink) exhibits a maximal Q factor, and a polarization singularity at the Γ point (Fig. 2(e)). From a polarization perspective, as the in-plane wavevector deviates from Γ point, this mode evolves into a high-$Q$ quasi-BIC with tangential polarization, indicating a vector light field. This polarization nature of the quasi-BIC is revealed by analyzing the momentum-space intensity distributions under different detection polarization angles $\theta$, defined as the in-plane angle between the transmission axis of the polarizer and the $k_x$ axis (Fig. 2(f)). At $\theta = 0°$, the $TE_{d1}$-derived BDL pattern exhibits bright lines parallel to $k_x$, while radiation along the $k_y$ axis nearly disappears. At $\theta = 90°$, the pattern shows double bright lines parallel to $k_y$. At $\theta = 45°$ and $135°$, orthogonal crossed bright lines appear, reflecting the tangential polarization nature of the quasi-BIC.

In the dual-BIC regime at $\alpha = -0.04$ in Fig. 2(b), two quasi-BICs corresponding to $TE_{d1}$ and $TM_{d1}$ (blue) are simultaneously excitable, each exhibiting its own characteristic far-field polarization (Fig. 2(e)). The $TE_{d1}$-derived quasi-BIC retains its tangential polarization, giving rise to the BDL pattern with the polarization behavior described above. The $TM_{d1}$-derived quasi-BIC exhibits a distinct radial polarization signature, which determines how its ring-shaped radiation pattern responds to $\theta$ (Fig. 2(f)), with bright spots that rotate with $\theta$ and always aligning with the polarizer axis. When superimposed with the $TE_{d1}$-derived BDL, the combined radiation pattern inherits both the tangential polarized BDL with and the ring with radial polarization. In the dual-BIC regime at α = –0.18 in Fig. 2(c), the simultaneously excitable quasi-BICs arise from $TE_{d1}$ and $TE_{d2}$. The $TE_{d1}$-derived quasi-BIC contributes a BDL pattern with tangential polarization, whereas the $TE_{d2}$-derived one exhibits an azimuthal polarization signature (Fig. 2(e)). Its ring-shaped radiation pattern produces bright spots that rotate with θ, appearing orthogonal to the polarization axis at $\theta = 0°$ and $90°$, but aligned with it at $\theta = 45°$ and $135°$ (Fig. 2(f)). This behavior contrasts with the radially polarized $TM_{d1}$-derived quasi-BIC, highlighting the diverse polarization characteristics accessible in the dual-BIC regime.

In the radiative mode with BIC regime at $\alpha = -0.23$ (Fig. 2(d)), the $TE_{d1}$-derived quasi-BIC and the radiative mode $TE_{b2}$ (red) can be simultaneously excited. The $TE_{d1}$-derived quasi-BIC again contributes a BDL pattern with tangential polarization. Unlike the quasi-BICs discussed above, $TE_{b2}$ is a radiative mode that radiates directly at the Γ point. Its field distribution exhibits a well-defined linear polarization oriented along the $k_x$ direction (Fig. 2(e)), producing a spot-like radiation pattern centered at the Γ point (Fig. 2(f)). When superimposed with the $TE_{d1}$-derived BDL, the combined pattern exhibits that the central spot appears only when $\theta$ aligns with its axis ($\theta = 0°$), vanishing at $\theta = 90°$, leaving only the BDL pattern with tangential polarization. As such, the controlled lasing patterns can be achieved by engineering $\alpha$, enabling distinct vectorial laser fields of tangentially polarized BDL, radially or azimuthally polarized rings, a linearly polarized spot, as well as the corresponding composite patterns. The vectorial lasing behavior across these regimes further enhances pattern controllability, offering a promising pathway for dynamic vectorial lasing.

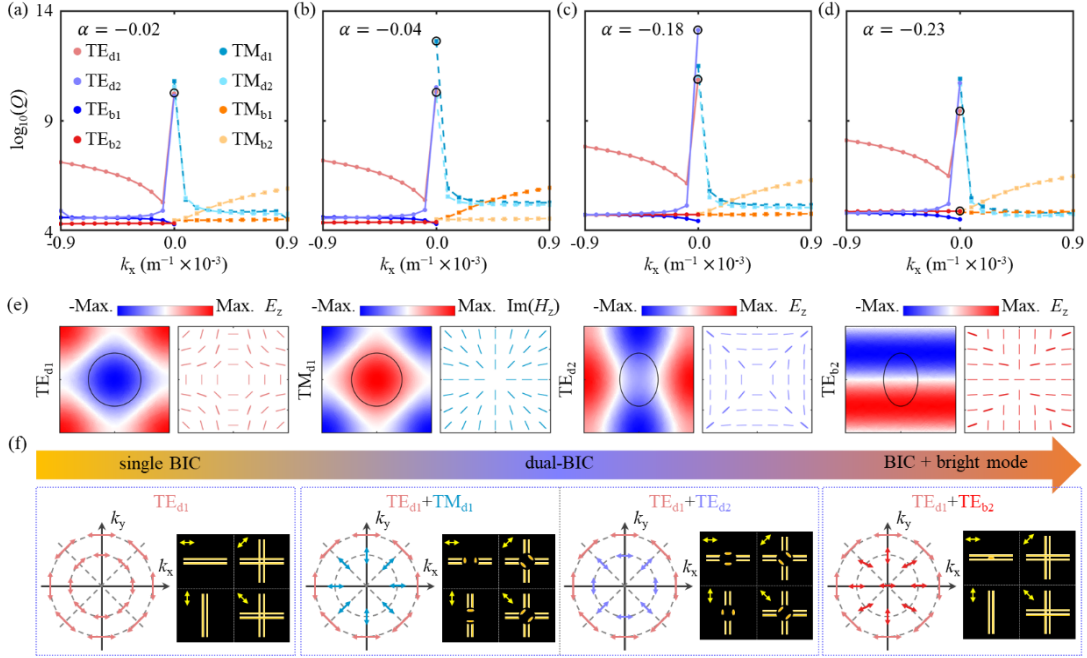

Fig. 2. Mode characteristics and polarization-controlled radiation patterns in the 2D q-BIC microcavity. (a–d) $Q$-factor evolution along Γ–X at representative $\alpha$ values corresponding to the three regimes: single BIC regime at $\alpha = -0.02$ (a), dual-BIC regime at $\alpha = -0.04$ (b) and $\alpha = -0.18$ (c), and radiative mode with BIC regime at $\alpha = -0.23$ (d). (e) Field distributions and far-field polarizations of key excitable modes. (f) Schematic momentum-space radiation patterns under various polarization angles $\theta$ for each $\alpha$, showing polarization-governed pattern characteristics.

**Realizing four type of vectoral lasing in momentum-space**

Based on the theoretical predictions, 2D quasi-BIC microcavities with varying $\alpha$ values were fabricated using dual-beam laser interference lithography with low-cost and high-throughput production (Fig. S3). The asymmetry factor $\alpha$ was precisely controlled by adjusting the exposure time during two orthogonal exposures. Taking the microcavity with $\alpha = -0.02$ as a representative example, scanning electron microscope (SEM) images confirm its well-defined periodic structure (Fig. S3(c)). All measurements were performed using a home-made angle-resolved spectra and momentum-space imaging system (Fig. S4).

To achieve the single quasi-BIC lasing, the microcavity with $\alpha = -0.02$ was first investigated, corresponding to the single BIC regime. Its dispersion properties were characterized through angle-resolved transmission spectra (Fig. S5). Resonant features are observed near the Γ point for both TE and TM waves, arising from the coupling of the first-order Bloch modes, confirming the presence of the expected resonant modes. Under optical pumping with a 532 nm nanosecond pulsed laser with the polarization angle $\theta_{in}$ fixed at 135° (green arrow in the inset of Fig. 3(a)), the device exhibits a broad spontaneous emission profile at the pump energy density of 0.386 mJ/cm². When the pump energy density slightly exceeds 0.395 mJ/cm², an extremely sharp emission peak emerges at 599.432 nm with a linewidth of 0.100 nm, corresponding to a $Q$ value of approximately 5994 (Fig. 3(a)). The integrated intensity and full width at half maximum (FWHM) as functions of pump energy density give a lasing threshold of 0.395 mJ/cm² (Fig. 3(b)). Above threshold, momentum-space imaging reveals a BDL pattern around the Γ point, characterized by the bright lines parallel to $k_x$ and $k_y$ axes with dark lines at $k_x = 0$ and $k_y = 0$ as theoretical analysis (inset of Fig. 3(a) and Fig. 3(c)). The absence of radiation at the Γ point in momentum space confirms that the lasing emission originates from a typical quasi-BIC from $TE_{d1}$[46]. Angle-resolved photoluminescence spectra further confirm the quasi-BIC nature. Below threshold, a grating diffraction pattern is observed (Fig. 3(d)), whereas above threshold, a distinct two-lobe pattern emerges, symmetrically distributed on either side of the Γ point (Fig. 3(e)), which is a characteristic radiation signature of quasi-BIC lasing[34]. Polarization-resolved momentum-space imaging at a pump energy density of 0.446 mJ/cm² (Fig. 3(f)) directly verifies the theoretical prediction in Fig. 2(f). At $\theta = 0°$, the BDL pattern exhibits strong intensity along the $k_x$ axis, while radiation along the $k_y$ axis nearly disappears. At $\theta = 90°$, the radiation intensity shifts to both sides of the $k_y$ axis. At $\theta = 45°$ and 135°, both the bright lines along the $k_x$ and $k_y$ axes are preserved, exhibiting a BDL pattern. This exact match confirms that the lasing originates from the $TE_{d1}$-derived quasi-BIC, and demonstrates that the radiation pattern can be switched among three distinct configurations simply by adjusting the polarization angle.

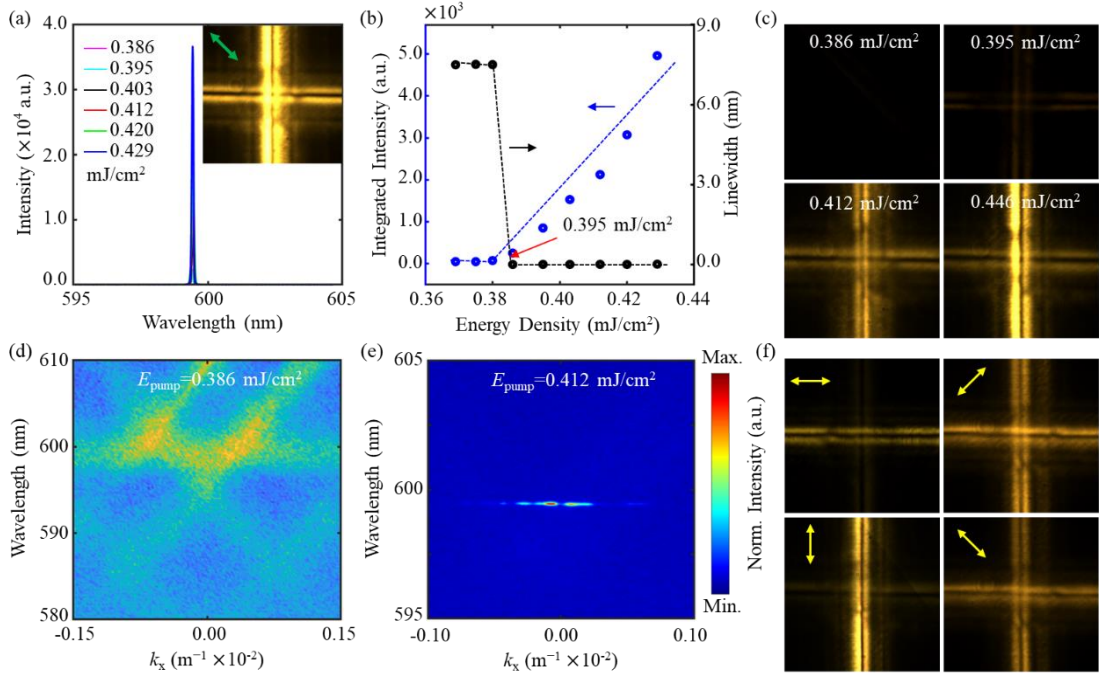

Fig. 3. Single quasi-BIC lasing from mode $TE_{d1}$ in the $\alpha = -0.02$ microcavity. (a) Lasing spectrum at different pump energy densities. Inset: momentum-space BDL pattern. (b) Integrated intensity and FWHM versus pump energy density. (c) Momentum-space image versus pump energy density, exhibiting a BDL pattern above threshold. (d-e) Angle-resolved photoluminescence spectrum below (d) and above threshold (e). (f) Polarization-resolved momentum-space images at 0.446 mJ/cm$^2$, revealing tangential polarization for BDL pattern.

To obtain a dual quasi-BIC lasing, the square PC quasi-BIC lasers with $\alpha = -0.04$ (Fig. S6(a)) was first fabricated to achieve the lasing modes from $TE_{d1}$ and $TM_{d1}$. The lasing threshold is measured to be 0.395 mJ/cm$^2$, and just above threshold a lasing peak is observed at 598.468 nm with the FWHM of 0.101 nm ($Q \approx 5925$) (Fig. 4(a) and Fig. S6(b)). As the pump energy density increases to 0.412 mJ/cm$^2$, additional narrow lasing peak emerges at 599.583 nm, with the linewidths less than 0.11 nm and $Q$ value near 5450. Angle-resolved photoluminescence spectra confirm that both lasing modes are quasi-BICs (Fig. S6(c)). Momentum-space imaging under varying pump energy density reveals the evolution between laser modes. At a low pump energy density of 0.395 mJ/cm$^2$, a symmetric ring-shaped pattern dominates, indicating that only $TM_{d1}$-derived quasi-BIC is excited (Fig. 4(b)). Polarization analysis confirms its radially polarized nature (Fig. S6(d)). At a higher pump energy density of 0.412 mJ/cm$^2$, the $TE_{d1}$-derived BDL pattern becomes visible and superimposes with the ring pattern (Fig. 4(b)), yielding a donut with BDL configuration characterized by a dark region between the two features, which might be attributed to mode competition between the orthogonally polarized contributing modes. Therefore, by simply varying the pump energy density, the dynamic reconfiguration of the radiation patterns between a single donut and a donut with BDL is achieved within the same device, as the two modes possess different lasing thresholds. Under polarization control at pump energy density of 0.452 mJ/cm$^2$ (Fig. 4(c)), the combined pattern exhibits the polarization-dependent behavior predicted in Fig. 2(f). At $\theta = 0°$, bright lines parallel to the $k_x$ axis from TE$_{d1}$ persist alongside a pair of bright spots from TM$_{d1}$ aligned along the $k_x$ axis. At $\theta = 90°$, bright lines parallel to the $k_y$ axis appear with a pair of bright spots aligned along the $k_y$ axis. At $\theta = 45°$ and $135°$, the complete BDL pattern remains visible while the spots orient along the diagonal axes.

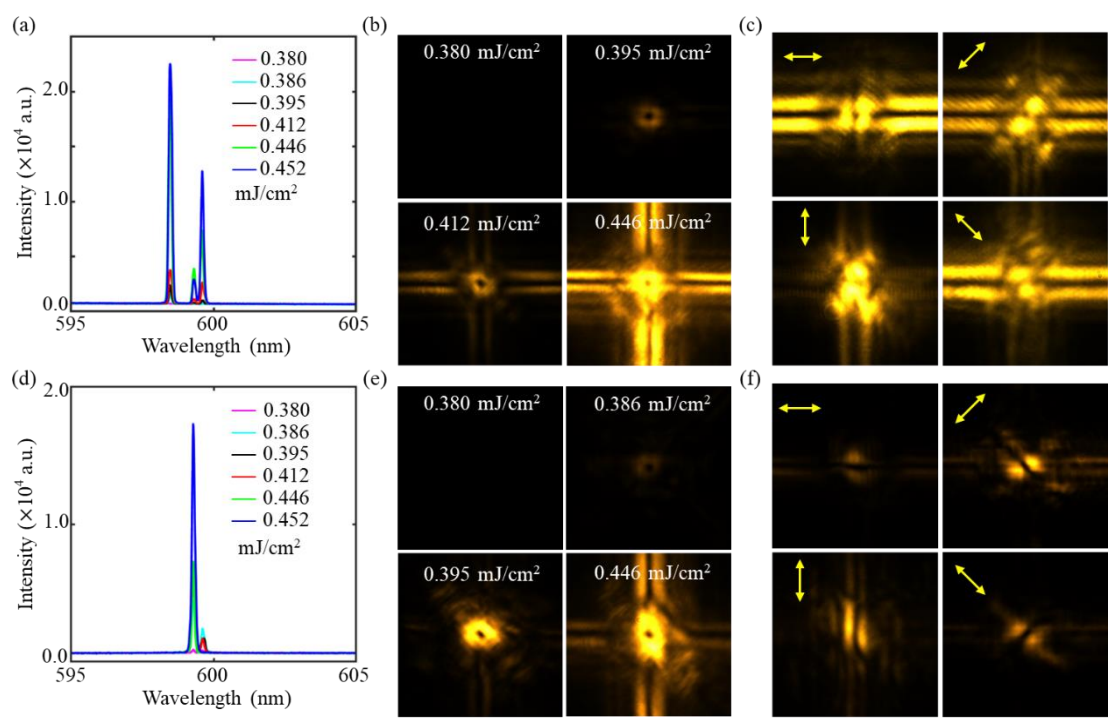

Fig. 4. Dual-BIC lasing with superimposed radiation patterns of BDL and ring. For $\alpha = -0.04$: (a) Lasing spectra at varying pump energy densities, showing peaks from modes TM$_{d1}$ (598.468 nm) and TE$_{d1}$ (599.583 nm). (b) Radiation patterns in momentum space under different pump energy densities. (c) Polarization-resolved momentum-space images at 0.446 mJ/cm$^2$, revealing radial polarization for ring. For $\alpha = -0.18$: (d) Lasing spectra at varying pump energy densities, showing peaks from modes TE$_{d2}$ (599.277 nm) and TE$_{d1}$ (599.603 nm). (e) Radiation patterns in momentum space under different pump energy densities. (f) Polarization-resolved momentum-space images at 0.446 mJ/cm$^2$, revealing azimuthal polarization for ring.

To obtain another dual quasi-BIC lasing, the square PC with $\alpha = -0.18$ (Fig. S7(a)) were further fabricated, exhibiting a lasing threshold of 0.386 mJ/cm² for the mode at 599.277 nm (Fig. 4(d) and Fig. S7(b)). The corresponding FWHM is about 0.104 nm ($Q \approx 5762$). Momentum-space radiation pattern evolves from a ring (Fig. 4(e)) corresponding to the $TE_{d2}$-derived quasi-BIC at lower pump energy to a combined pattern featuring both the ring and an additional BDL from the $TE_{d1}$-derived quasi-BIC at 599.603 nm as the pump energy density increases to 0.395 mJ/cm². Similar to the device with $\alpha = -0.04$, the reversible switching capability of the radiation patterns is also retained by varying the pump energy density. Angle-resolved photoluminescence spectra show that both modes generate symmetric two-lobe lasing condensates on either side of the Γ point (Fig. S7(c)). Polarization analysis of the ring-shaped pattern from $TE_{d2}$ reveals distinct behavior compared to that from $TM_{d1}$ (Fig. 4(f)). At $\theta = 0°$, bright spots appear along the $k_y$ axis. At $\theta = 90°$, the spots shift to the $k_x$ axis. At $\theta = 45°$ and 135°, the spots emerge along the corresponding 45° or 135° directions. This orthogonal relationship between spot orientation and polarization angle directly reflects the distinct polarization signature of $TE_{d2}$-derived quasi-BIC, consistent with the theoretical predictions in Fig. 2(f).

To construct a radiative mode with quasi-BIC lasing, the 2D PC was designed at $\alpha = -0.23$ (Fig. S8(a)). Under a pump threshold of 0.395 mJ/cm², a narrow-linewidth resonant peak appears at 598.921 nm (Fig. 5(a) and Fig. S8(b)). When the pump energy density increases to 0.412 mJ/cm², the measured linewidth of this resonant peak is approximately 0.11 nm, exhibiting a $TE_{d1}$-derived BDL radiation pattern in momentum space (Fig. 5(b)). At a higher pump energy density of 0.446 mJ/cm², a spot-like radiation pattern from radiative mode $TE_{b2}$ emerges at the Γ point while the BDL pattern persists (Fig. 5(b)). The spectral linewidth broadens to 0.315 nm ($Q \approx 1900$), indicating the collective contribution of multiple resonant modes. Angle-resolved photoluminescence spectra reveal a bright spot at the Γ point (Fig. 5(c)), confirming the onset of Bragg resonance lasing[47]. Polarization analysis at 0.461 mJ/cm² reveals the distinct nature of the two coexisting modes (Fig. 5(d)). At $\theta = 0°$, the central spot-like pattern reaches maximum intensity while the BDL pattern shows radiation along the $k_x$ direction. At $\theta = 90°$, the central spot completely disappears, leaving only bright lines parallel to the $k_y$ axis. This polarization behavior demonstrates that the spot-like Bragg resonance lasing exhibits uniform linear polarization along the $k_x$ direction, whereas the $TE_{d1}$-derived quasi-BIC mode retains its characteristic radiation pattern.

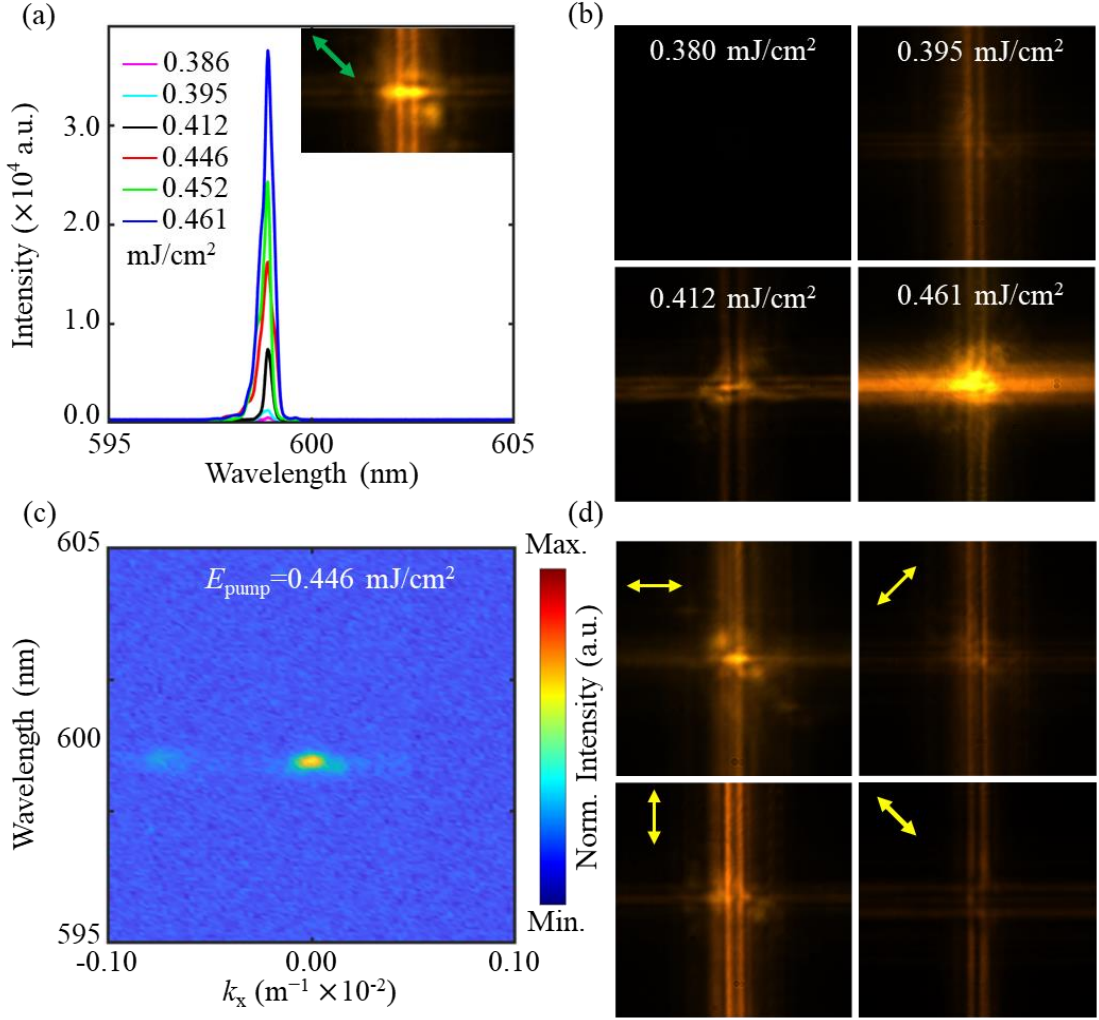

Fig. 5. Coexistence of quasi-BIC and Bragg resonance lasing in the $\alpha = -0.23$ microcavity. (a) Lasing spectra at varying pump energy densities, showing the transition from a narrow peak to a broadened spectral profile. (b) Momentum-space images at increasing pump energy densities: BDL pattern from $TE_{d1}$-derived mode at 0.412 mJ/cm², and superposition with spot-like pattern at Γ point at 0.446 mJ/cm². (c) Angle-resolved photoluminescence spectra at 0.446 mJ/cm², revealing bright spot at Γ point indicative of Bragg resonance lasing. (d) Polarization-resolved momentum-space images at 0.461 mJ/cm², showing linear polarization for central spot.

**Conclusion and discussions**

Based on 2D quasi-BIC lasers, we theoretically predict and experimentally demonstrate selective excitation of different quasi-BIC and radiative modes by engineering the geometric asymmetry factor $\alpha$, establishing a direct and versatile approach to control the number, type, and momentum-space polarization topology of lasing modes. By varying α, three distinct radiation regimes are identified, yielding four characteristic vectorial radiation patterns in momentum space. In the single BIC regime ($\alpha = -0.02$), lasing arises solely from the TE$_{d1}$-derived quasi-BIC, producing a tangentially polarized BDL pattern with tangential polarization. In the dual-BIC regime, multiple quasi-BICs are simultaneously excited, generating a ring pattern superimposed on a tangential-polarized BDL. The ring exhibits either radial polarization at $\alpha = -0.04$) or azimuthal polarization at $\alpha = -0.18$, representing two distinct vectorial outputs within the same regime. By varying the pump energy density, reversible switching between a pure donut and a donut with BDL pattern is achieved in the same device, enabling dynamic reconfigurability of momentum-space radiation profiles. In the regime of radiative mode with BIC at $\alpha = -0.23$, we realize a hybrid lasing state combining quasi-BIC lasing and Bragg resonance lasing, which yielding a linearly polarized spot at the Γ point superimposed on a tangential-polarized BDL. Moreover, polarization control further extends the ability to tailor momentum-space radiation patterns, adding an extra degree of freedom for shaping vectorial lasing fields. Importantly, the proposed approach of engineering the geometric asymmetry factor relies solely on structural parameters, making it readily transferable to other material platforms and spectral regimes. Collectively, these results establish that asymmetry factor engineering provides a powerful paradigm for controlling the complexity and polarization diversity of momentum-space vectorial lasing in 2D quasi-BIC microcavities. The ability to dynamically reconfigure radiation patterns and polarization states within a single device offers a compact and versatile platform for advanced photonic applications, including structured light generation, optical trapping, high-capacity communications, and on-chip vector beam control.

**Method**

**Theoretical model for 2D PC-waveguide-coupled structure**

To elucidate the interaction mechanisms between the resonance modes in a 2D PC and waveguide coupled structure, a generalized Bragg resonance theory is given to analyze the dispersion relation without the coupling.

For a PC-waveguide-coupled structure (Fig. S1(a)), the refractive index of the polymer is denoted as $n_{WG}$, the thickness of the waveguide is $T_{WG}$, and the lattice constant along the $x$ axial and $y$ axial of the PC are $\varLambda_x$ and $\varLambda_y$. In the PC-waveguide-coupled structure, the electromagnetic modes are quantized into different band branches named as the $TE_{l,p}^{m}$ for TE polarized waves and $TM_{l,p}^{m}$ for TM polarized waves due to the in-plane period boundary of the PC and the localization of the waveguide along $z$ direction. $l$ and $p$ indicate the diffraction orders induced by the PC through the Bloch condition along $x$ direction ($l = \pm1, \pm2, ......$) and $y$ direction ($p = \pm1, \pm2, ......$). And the waveguide further discretizes the $l$-th order and $p$-th order diffraction branch into several branches with $m$ representing the resonance order of the waveguide structure ($m$ = 0, 1, 2, ……) along $z$ direction.

For the modes of $TE_{l,p}^{m}$ (or $TM_{l,p}^{m}$), the incident wave vector $\vec{k}_{in}$ propagating at an angle $\varphi_{in}$ relative to the $z$-axis in the waveguide, its $z$-component in the waveguide satisfies the resonance condition:

$$\vec{k}_{\perp in} = \frac{m\pi + \Delta\varPhi}{T}, \quad (1)$$

where, $\Delta\varPhi$ denotes the negligible phase difference. The in-plane components of the wave vector $\vec{k}_{in}$ in the XY plane satisfy the interface momentum matching condition:

$$\vec{k}_{\|in} - \left(l\vec{G}_x + p\vec{G}_y\right) = \vec{k}_{\|out}. \quad (2)$$

Here, $\vec{k}_{\|in}$ denotes the in-plane wave vector component of the incident light in the waveguide, which satisfies $\vec{k}_{\|in} = \sqrt{\vec{k}_{in}^{\,2} - \vec{k}_{\perp in}^{\,2}} = n_{WG} \cdot k_0 \sin\varphi_{in}$. $\vec{G}_x$ ($\vec{G}_y$) represents the PC wave vector, which is periodic along the $x$ ($y$) direction, satisfying the relationship $\vec{G}_x = \frac{2\pi}{\varLambda_x}\hat{x}$ ($\vec{G}_y = \frac{2\pi}{\varLambda_y}\hat{y}$). $\vec{k}_{\|out}$ is the in-plane wave vector component of the emitted mode, satisfying $\vec{k}_{\parallel out} = k_0 \sin\varphi_{out}$ with the irradiation angle $\varphi_{out} \in \left(-\frac{\pi}{2}, \frac{\pi}{2}\right)$ in air. $k_0 = \frac{2\pi}{\lambda_0}$ is the wave vector in free space.

Accounting for the waveguide thickness $T_{WG}$, the relationship between the free-space wavelength $\lambda_0$ and the resonance modes $TE_{l,p}^m$ (or $TM_{l,p}^m$) is governed by:

$$\left(n_{WG} \cdot \frac{2\pi}{\lambda_0}\right)^2 - \left(\frac{m\pi}{T_{WG}}\right)^2 = \left(k_x - l\frac{2\pi}{\Lambda_x}\right)^2 + \left(k_y - p\frac{2\pi}{\Lambda_y}\right)^2 \tag{3}$$

This formula supplies the dispersion relation of the 2D PC-waveguide-coupled structure, indicating the origin of each band branch $TE_{l,p}^m$ (or $TM_{l,p}^m$) from the $l$-th ($p$-th) diffraction order and the $m$-th Bloch modes.

**Full-wave simulation**

In the 2D PC-waveguide-coupled structure, R6G-doped photoresist material ($n$=1.64) was used for both the PC and waveguide part on glass ($n$=1.5) substrate to form a quasi-BIC microcavity with a low index contrast. The PC layer thickness ($T_{PC}$) is set at 150 nm, and the lattice constants in the $x$ ($\Lambda_x$) and $y$ ($\Lambda_y$) directions are both 375 nm. According pillar axis lengths along the $x$ ($\Lambda_x$) and $y$ ($\Lambda_y$) directions, the geometric asymmetry factor is defined as $\alpha = (w_x - w_y)/(w_x + w_y)$, serving as key parameters for controlling mode coupling characteristics. The waveguide thickness ($T_{WG}$) represents another essential parameter that directly influences the dispersion properties of Bloch modes.

For numerical simulations of PC-waveguide-coupled structure, three-dimensional full-wave simulations using the finite-element method is employed. In the 3D modeling framework, periodic boundary conditions are imposed along the $x$ and $y$ directions. To calculate eigenmodes, perfectly matched layers are applied in the $z$ direction for effective radiation absorption. The mode analysis modules are utilized to solve for dispersion relations, electromagnetic field distributions, and $Q$ factors of the PC-waveguide-coupled structure. The $Q$ factor is determined by $Q$=(Re[$f$])/(2Im[$f$]), where Re[$f$] and Im[$f$] denote the real and imaginary parts of the eigenfrequency of a simulated cavity eigenmode, respectively. The research examines the influence of $\alpha$ variations within the range of −0.33 to 0.05 on the $Q$ factors of coupled modes at the fixed conditions of $\Lambda$=375 nm and $T_{WG}$=1.75 μm.

**Fabrication and Optical Measurements**

The photoresist (AZ® MiR™ 701 Series) films doped with R6G were deposited on the glass substrate through the spin-coating method. The 2D PC was then fabricated on surface of the film by using a dual-beam interference lithography technique. A continuous-wave laser (He-Cd Laser, CW, 325 nm) was employed as the interference light source. The geometric asymmetry factor of the PC was tuned by controlling the exposure time.

All optical measurements were performed using a home-made angle-resolved spectra and momentum-space imaging system designed with special optical lens configurations (Fig. S4, Supplementary Information). The system features exceptional angular resolution capability, achieving a numerical aperture (NA) of 0.22. A nanosecond laser (MCA-532-1-60-01-PD) was employed as the pump source, with a repetition rate of 500 Hz and a central wavelength of 532 nm. The pump energy density was controlled by a half-wave plate and a linear polarizer. The polarization state of the emitted light was analyzed using a linear polarizer. The angle-resolved transmission and photoluminescence spectra were obtained by using a spectrometer (Andor-SR-500i-D1-R) with a 2D CCD array at back focal plane. The momentum-space images were recorded by the camera (RH-CAM4K8MPA).

**Supporting Information**

The Supporting Information is available free of charge via the journal's website.

**Acknowledgments**. National Natural Science Foundation of China (grant no. 92150109 and 61975018).

**Disclosures**. The authors declare no conflicts of interest.

**Data availability**. Data underlying the results presented in this paper are not publicly available at this time but may be obtained from the authors upon reasonable request.

**References:**
[1] Liu W. Z., Shi L., Chan C. T., et al. Momentum-space polarization fields in two-dimensional photonic-crystal slabs: Physics and applications. *Chinese Physics B* **31**, 104211 (2022).
[2] Che Z. y., Zhang Y. b., Liu W. z., et al. Polarization Singularities of Photonic Quasicrystals in Momentum Space. *Physical Review Letters* **127**, 043901 (2021).
[3] Wang B., Liu W. z., Zhao M. x., et al. Generating optical vortex beams by momentum-space polarization vortices centred at bound states in the continuum. *Nature Photonics* **14**, 623-628 (2020).
[4] Wang W. M., Kou J. L., and Lu Y. Q. Polarization Field in Momentum Space of Two-Dimensional Photonic Crystal Slabs. *Acta Optica Sinica* **44**, 1026003 (2024).
[5] Hu Z. X., Li K. F., Li C., et al. Spin–Orbit Interaction Enabled Nonlinear Metasurface Holography. *Advanced Materials* **38**, e21229 (2026).
[6] Padgett M., and Bowman R. Tweezers with a twist. *Nature Photonics* **5**, 343-348 (2011).
[7] Ahmed H., Ansari M. A., Paterson L., et al. Vector Vortex Beam-Enabled Edge Microscopy with Dynamic Orientation Selectivity. *ACS Photonics* **12**, 7013-7019 (2025).
[8] Forbes A., de Oliveira M., and Dennis M. R. Structured light. *Nature Photonics* **15**, 253-262 (2021).
[9] Angelsky O. V., Bekshaev A. Y., Mokhun I. I., et al. Review on the structured light properties: rotational features and singularities. *Opto-Electronics Review* **30**, 140860-140860 (2022).
[10] Zhu Z., Janasik M., Fyffe A., et al. Compensation-free high-dimensional free-space optical communication using turbulence-resilient vector beams. *Nature Communications* **12**, 1666 (2021).
[11] Naidoo D., Roux F. S., Dudley A., et al. Controlled generation of higher-order Poincaré sphere beams from a laser. *Nature Photonics* **10**, 327-332 (2016).
[12] Chen Y., Wang M. J., Si J. H., et al. Observation of chiral emission enabled by collective guided resonances. *Nature Nanotechnology* **20**, 1205-1212 (2025).
[13] Pan Fu, Pei Nan Ni, Bo Wu, et al. Metasurface Enabled On‐Chip Generation and Manipulation of Vector Beams from Vertical. *Advanced Materials* **35**, 2204286 (2023).
[14] Zhai Z. S., Li Z., Du Y. X., et al. Multimode Vortex Lasing from Dye–TiO2 Lattices via Bound States in the Continuum. *ACS Photonics* **10**, 437-446 (2023).
[15] Xing D., Chen M. H., Wang Z. Y., et al. Solution‐Processed Perovskite Quantum Dot Quasi‐BIC Laser from Miniaturized Low‐Lateral‐Loss Cavity. *Advanced Functional Materials* **34**, 2314953 (2024).
[16] Miao P., Zhang Z. f., Sun J. b., et al. Orbital angular momentum microlaser. *Science* **353**, 464-467 (2016).
[17] Hu P., Wang J. J., Jiang Q., et al. Global phase diagram of bound states in the continuum. *Optica* **9**, 1353-1361 (2022).
[18] Liu W. Z., Wang B., Zhang Y. W., et al. Circularly Polarized States Spawning from Bound States in the Continuum. *Physical Review Letters* **123**, 116104 (2019).
[19] Doeleman H. M., Monticone F., den Hollander W., et al. Experimental observation of a polarization vortex at an optical bound state in the continuum. *Nature Photonics* **12**, 397-401 (2018).
[20] Wu J. J., Chen J. G., Qi X., et al. Observation of accurately designed bound states in the continuum in momentum space. *Photonics Research* **12**, 638-647 (2024).
[21] Zhuang Z. P., Zeng H. L., Chen X. D., et al. Topological Nature of Radiation Asymmetry in Bilayer Metagratings. *Physical Review Letters* **132**, 113801 (2024).
[22] Jiang Q., Hu P., Wang J., et al. General Bound States in the Continuum in Momentum Space. *Physical Review Letters* **131**, 013801 (2023).
[23] Zhang X. D., Liu Y. L., Han J. C., et al. Chiral emission from resonant metasurfaces. *Science* **377**, 1215–1218 (2022).
[24] Zhou X. B., Li Z., Zhou Y. F., et al. Lasing from Doubly Degenerate Bound States in the Continuum. *The Journal of Physical Chemistry Letters* **15**, 10703-10709 (2024).
[25] Xu Z. Y., Liu Y., Chang S. Q., et al. Ultrathin Deployable Femtosecond Vortex

Laser. *Advanced Materials* **37**, 2507122 (2025).
[26] Hwang M. S., Lee H. C., Kim K. H., et al. Ultralow-threshold laser using super-bound states in the continuum. *Nature Communications* **12**, 4135 (2021).
[27] Ren Y. H., Li P. S., Liu Z. J., et al. Low-threshold nanolasers based on miniaturized bound states in the continuum. *Science Advances* **8**, eade8817 (2022).
[28] Zhong H. C., Yu Y., Zheng Z. Y., et al. Ultra-low threshold continuous-wave quantum dot mini-BIC lasers. *Light: Science & Applications* **12**, 100 (2023).
[29] Wu X. X., Zhang S., Song J. P., et al. Exciton polariton condensation from bound states in the continuum at room temperature. *Nature Communications* **15**, 3345 (2024).
[30] Guan J., Sagar L. K., Li R., et al. Quantum Dot-Plasmon Lasing with Controlled Polarization Patterns. *ACS Nano* **14**, 3426-3433 (2020).
[31] Tang H. J., Huang C., Wang Y. H., et al. Dynamically tunable long-range coupling enabled by bound state in the continuum. *Light: Science & Applications* **14**, 278 (2025).
[32] Wang Y. H., Fan Y. B., Zhang X. D., et al. Highly Controllable Etchless Perovskite Microlasers Based on Bound States in the Continuum. *ACS Nano* **15**, 7386-7391 (2021).
[33] Wu M. f., Ha S. T., Shendre S., et al. Room-Temperature Lasing in Colloidal Nanoplatelets via Mie-Resonant Bound States in the Continuum. *Nano Letters* **20**, 6005-6011 (2020).
[34] Do T. T. H., Yuan Z. Y., Durmusoglu E. G., et al. Room-Temperature Lasing at Flatband Bound States in the Continuum. *ACS Nano* **19**, 19287-19296 (2025).
[35] Wang M. J., Lv N. Y., Zhang Z. X., et al. Chiral orbital lasing in a twisted bilayer metasurface. *Nature Communications* **17**, (2026).
[36] Zhang T. C., Dong K. C., Li J. C., et al. Twisted moiré photonic crystal enabled optical vortex generation through bound states in the continuum. *Nature Communications* **14**, 6014 (2023).
[37] Zeng Y. X., Sha X. B., Zhang C., et al. Metalasers with arbitrarily shaped wavefront. *Nature* **643**, 1240-1245 (2025).
[38] Qin H. Y., Su Z. P., Zhang Z., et al. Disorder-assisted real-momentum topological photonic crystal. *Nature* **639**, (2025).
[39] Chai R. H., Liu W. W., Li Z. C., et al. Spatial Information Lasing Enabled by Full-k-Space Bound States in the Continuum. *Physical Review Letters* **132**, (2024).
[40] Yuan H. Y., Liu J. Y., Wang X. L., et al. Dynamically Switchable Polarization Lasing Between q‐BIC and Bragg Resonance Modes. *Laser & Photonics Reviews* **20**, e01175 (2025).
[41] Dai S. W., Hu P., and Han D. Z. Near-field analysis of bound states in the continuum in photonic crystal slabs. *Optics Express* **28**, 16288-16297 (2020).
[42] Weber T., Kühner L., Sortino L., et al. Intrinsic strong light-matter coupling with self-hybridized bound states in the continuum in van der Waals metasurfaces. *Nature Materials* **22**, 970-976 (2023).
[43] Ardizzone V., Riminucci F., Zanotti S., et al. Polariton Bose–Einstein condensate from a bound state in the continuum. *Nature* **605**, 447-452 (2022).
[44] Hsu C. W., Zhen B., Stone A. D., et al. Bound states in the continuum. *Nature Reviews Materials* **1**, 16048 (2016).
[45] Koshelev K., Sadrieva Z., Shcherbakov A., et al. Bound states in the continuum in photonic structures. *Physics Uspekhi* **66**, 494-517 (2022).
[46] Ha S. T., Paniagua‐Domínguez R., and Kuznetsov A. I. Room‐Temperature Multi‐Beam, Multi‐Wavelength Bound States in the Continuum Laser. *Advanced Optical Materials* **10**, 2200753 (2022).
[47] Hakala T. K., Rekola H. T., Väkeväinen A. I., et al. Lasing in dark and bright modes of a finite-sized plasmonic lattice. *Nature Communications* **8**, 13687 (2017).